\documentclass[%
reprint,
 amsmath,amssymb,
 aps,
 prb,
]{revtex4-1}

\usepackage{graphicx}
\usepackage{bm}

\begin{document}

\title{Effect of quasi-bound states on coherent electron transport in twisted nanowires}

\author{Giampaolo Cuoghi}
\affiliation{Centro S3, CNR - Istituto Nanoscienze, Via Campi 213/A, 41100
Modena, Italy}
\affiliation{Dipartimento di Fisica,
Universit\`a degli Studi di Modena e Reggio Emilia,
Modena, Italy}

\author{Andrea Bertoni}
\email[e-mail: ]{andrea.bertoni@unimore.it}
\affiliation{Centro S3, CNR - Istituto Nanoscienze, Via Campi 213/A, 41100
Modena, Italy}

\author{Andrea Sacchetti}
\affiliation{Dipartimento di Matematica Pura ed Applicata,
Universit\`a degli Studi di Modena e Reggio Emilia,
Modena, Italy}

\date{\today}

\begin{abstract}
Quantum transmission spectra of a twisted electron waveguide expose
the coupling between traveling and quasi-bound states. Through a
direct numerical solution of the open-boundary Schr\"odinger equation
we single out the effects of the twist and show how the presence of a
localized state leads to a Breit-Wigner or a Fano resonance in the
transmission.  We also find that the energy of quasi-bound states is
increased by the twist, in spite of the constant section area along
the waveguide.  While the mixing of different transmission channels is
expected to reduce the conductance, the shift of localized levels into
the traveling-states energy range can reduce their detrimental effects
on coherent transport.
\end{abstract}



\maketitle

\section{Introduction}

Conductance spectra of quasi-1D semiconductor structures display many
features that expose directly the quantum nature of carrier transport,
and are of great interest both for applications and fundamental
understandings\cite{ihnbook04,rurali10}.  Even in the simplest
non-interacting carriers approach, the departure from a constant
section of the wire gives rise to complex resonance patterns in the
quantum transmission.  This originates from the coherent coupling of
the energy spectra of different subbands and from the interplay of
traveling and localized states\cite{nockel94}.  Indeed, the case of a
discrete energy spectrum merged with a continuum one, was considered
by Fano\cite{fano61} in his seminal work on inelastic scattering
amplitudes of electrons.  In that case, the two Hamiltonians with
discrete and continuous spectra were that of the electronic degree of
freedom of an atom and that of a free electron, respectively.  It was
shown that the coupling induced by the Coulomb interaction led to a
peculiar asymmetric shape of the scattering probability and a
discontinuity of the scattering phase.  This behavior of the
scattering amplitude is now identified in many
atomic\cite{fransson07}, optical\cite{rotter04} and
transport\cite{gores00} experiments (for a review see
Ref.~\onlinecite{miroshnichenko10}).

\begin{figure}[b]
\includegraphics[width=\linewidth,trim=19mm 0 0 10mm, clip=true]{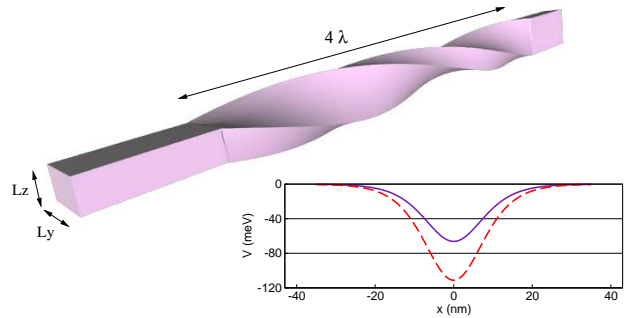}
\caption{
Quantum waveguide with rectangular cross-section ($L_y=20$~nm,
$L_z=10$~nm) twisted for a length of $4\lambda \simeq 70$~nm.  Here the
rotation angle is $\Phi=\frac{3}{2}\pi$.  Inset: local confining
potential, as given in Eq.~(\ref{eq2}), with $L_p= 10$~nm and
$\nu=2.95$ (solid line) or $\nu=3.95$ (dashed line). The potential
well is fully contained in the twisted region.  }\label{fig1}
\end{figure}

Here, we analyze the coherent transmission of a quantum waveguide (QW)
locally twisted, as depicted in Fig.~\ref{fig1}, with the twist
inducing a coupling between the subbands related to different
transverse modes.
A local attractive potential is also included, in order to give rise
to a discrete set of bound states and to Fano resonances in the
transmission spectra: they will expose the energy of quasi-bound
states of the twisted QW.  We stress that our results are
representative of a more general case, as for example a carrier
scattered through a quantum dot embedded in a QW or a QW whose
Hamiltonian is not separable in the transverse and longitudinal
directions, leading to localized states.

The effect of twisting on the conductance is twofold.  On one side, it
is expected to reduce the conductance\cite{Suzuura, Tseng}.  On the
other side, this reduction can be compensated by a partial destruction
of localization effects (e.g. due to external fields or impurities),
induced by the twisting itself.
In fact, by means of the complex scaling method it has been
shown\cite{kovarik07} that stable states associated to a trapping
potential may become resonant states when the QW is twisted.  In this
paper we numerically compute the real and imaginary part of such
resonances as a function of the twisting parameter in an explicit
model and we prove that, for such a model, the imaginary part of these
resonances actually takes a negative value, this indicating that the
corresponding states become unstable.
Specifically, our results show that the energy of bound states of the
quantum well in the longitudinal direction is increased by the twist
and, as they enter the continuous-spectrum range of traveling states,
they appear in the the transmission characteristic as symmetric
(Breit-Wigner~\cite{breit36}) or asymmetric (Fano~\cite{fano35})
resonant peaks, according to the character of the original bound wave
function.  The width of the resonances is related to the imaginary
component of the eigenvalues of the complex-scaled
Hamiltonian~\cite{reinhardt82}, as we detail in the following, and
shows a non-monotonic behavior. A resonant peak may or may not
disappear when its energy reaches the transmission channel with the
same transverse energy as the original bound state, according to
the corresponding bound state in the straight QW. Indeed,
the knowledge of the bound states of the straight QW allows one to
predict the position and type of transmission resonances in the
twisted system.

Our work is organized as follows. In Sect.~\ref{sec2} we describe the
model of the twisted QW and, in the following Sect.~\ref{sec3}, we
outline the real-space numerical approach adopted for the calculation
of the scattering states and transmission amplitudes on a
non-Cartesian grid.  In Sect.~\ref{sec4} the main results of our study
are presented, with particular attention to the evolution of resonant
peaks with the QW twist.  Finally, in Sect.~\ref{sec5} we draw our
conclusions. In the final Appendix, analytical details of the
complex-scaling approach mentioned in the main text, are given.

\section{The physical system}
\label{sec2}

We consider a QW with rectangular cross-section, with an hard-wall
confinement.  For a straight wire, an elementary solution of the
single-band effective-mass Schr\"odinger equation gives the energy
spectrum
\begin{equation}
E_{n,k}= E_n + \frac{\hbar^{2}}{2m}k^2 ,
\end{equation}
with
\begin{equation}  \label{eq1}
E_n=\frac{\hbar^{2}}{2m}\left[
\left(\frac{n_y\pi}{L_{y}}\right)^{2}+
\left(\frac{n_z\pi}{L_{z}}\right)^{2}\right] ,
\end{equation}
where $m$ is the effective mass of the carrier, $L_y$ and $L_z$
(with $L_y \neq L_z$) are
the thicknesses of the QW in the two directions orthogonal to the
current propagation, $k$ is the wave number of the $x$-propagating
plane wave component of the wave function. For the sake of brevity,
the subband index $n= 1,2,\dots$ (with $E_{n,k} \leq E_{(n+1),k}$)
has been introduced,
summarizing the two positive integers $n_y$ and $n_z$.
Since $k$ can be any real number, it is clear that the energy spectrum
is continuum, with $E_{n,k} \in [E_{1}\, ,+\infty)$.

A confining potential well (depicted in Fig.~\ref{fig1} inset)
is introduced along the $x$ direction
\begin{equation}  \label{eq2}
V\left(x\right)=\frac{\hbar^{2}}{2m}\frac{\nu\left(\nu+1\right)}
{L_{p}^{2}}\left[\tanh^{2}\left(\frac{x}{L_{p}}\right)-1\right],
\end{equation}
where the positive parameters $\nu$ and $L_p$ set the depth and the
length of the well. Specifically, the minimum of $V$ is
$-\hbar^{2}\nu\left(\nu+1\right)/(2mL_{p}^{2})$, and the region in
which $V$ is significantly different from zero is about $6\,L_p$. We
stress that the form given in Eq.~(\ref{eq2}) has been chosen both to
mimic a ``smooth'' local confinement and to deal with a potential that
has an exact expression for its bound-state eigenvalues~\cite{morsebook53}:
\begin{equation} \label{eq3}
\mu_{j} = -\frac{\hbar^{2}}{2mL_{p}^{2}}\left(\nu+1-j\right)^{2},
\; j=1,2,\ldots,\lceil \nu  \rceil \, ,
\end{equation}
where the ceiling function $\lceil \nu \rceil$ indicates the smallest
integer not less than $\nu$.  The above expression is essential to
approach analytically the problem through the complex scaling method,
that we use to follow the energy vs. twist behavior of the bound
states.

The energetic spectrum of the Schr\"odinger operator for the QW with
$V$ consists of a discrete and a continuum part
\begin{equation} \label{eq4}
\left\{ E_{n}+\mu_{j} ; \, n=1,2,\ldots ; 
\, j=1,\ldots, \lceil \nu  \rceil \right\} 
\cup [E_{1}\, , +\infty ) .
\end{equation}
For energies above $E_{1}$ the two parts overlap and eigenvalues of
the discrete spectrum are embedded in the continuum spectrum.
However, the two sub-spectra remain well distinct since the system
Hamiltonian is separable in a transverse ($y$-$z$ plane) and a
longitudinal ($x$ direction) component.  In fact, if a given energy
corresponds to a discrete level and, at the same time, it lies inside
the continuum, the corresponding state will be degenerate, with the
different eigenfunctions having a different transverse state.  In
terms of quantum transport along the QW, the above system does not mix
different transmission channels or propagating states with bound ones.

Let us now introduce a local twist in the QW. As we will show, this
couples different transverse modes, mixing their spectra, and shifts
the energy of the discrete states.  The deformation adopted, also
illustrated in Fig.~\ref{fig1}, is a rotation of the rectangular
cross-section around its center when moving along the QW axis.  A
point of coordinates $(x,y,z)$ of the straight QW is transformed
according
\begin{eqnarray} \label{eq5}
x' &=& x ,  \nonumber \\
y' &=& y\cos\left[\varphi\left(x\right)\right]
+z\sin\left[\varphi\left(x\right)\right],
   \\
z' &=& z\cos\left[\varphi\left(x\right)\right]
-y\sin\left[\varphi\left(x\right)\right] ,
\nonumber
\end{eqnarray}
where
\begin{equation} \label{eq6}
\varphi\left(x\right) = 
\frac{\Phi}{2}\left[\textrm{erf}\left(\frac{x}{\lambda}\right)+1\right]
\end{equation}
is the rotation angle as a function of the longitudinal position $x$.
Here, $\textrm{erf}$ is the error function, $\Phi$ is the total
rotation angle, and $\lambda$ is a parameter that sets the length of
the twisted region. In particular, the QW twist can be considered
effective in a length $4 \, \lambda$ around the origin. Outside the
latter region, the QW is essentially straight.

\begin{figure}
\includegraphics[width=.7\linewidth]{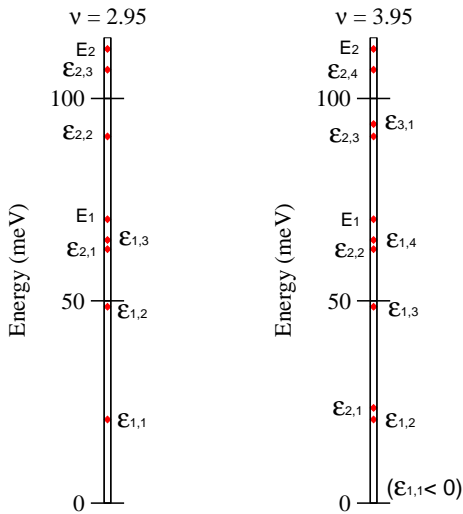}
\caption{
Schematic representation of the relevant energies of the system
without twist for two different magnitudes of the quantum well
described by Eq.~(\ref{eq2}).  The thresholds of the transport
channels, i.e. the transverse modes, are $E_1=70.155$~meV,
$E_2=112.248$~meV, $E_3=182.403$~meV (not shown).  The energy of a
bound state of the straight QW is indicated with $\varepsilon_{n,j}$,
with $n$ indicating the transverse mode, and $j$ the bound state of
$V$ in the longitudinal direction.  While the $E_n$'s are fixed, the
$\varepsilon_{n,j}$'s are increased by the twist.  }\label{fig2}
\end{figure}

For our simulations, we use GaAs effective electron mass
$m=0.067\,m_e$ and adopt the following set of geometric parameters:
$L_{y}=20$~nm, $L_{z}=10$~nm, $\lambda=17.5$~nm (i.e. the twisted
region is about $70$~nm), $0\leq\Phi\leq 3\pi$.  Two attractive
potentials, as given in Eq.~(\ref{eq2}), have been used, both with
$L_{p}=10$~nm (i.e. effective on a length of about $60$~nm around the
origin).  They differ by their depth: in the first case $\nu=2.95$
(corresponding to a minimum of $-66.262$~meV), in the second case
$\nu=3.95$ (corresponding to a minimum of $-111.186$~meV). The
relative positions of relevant energy levels are reported in
Fig.~\ref{fig2}, for the two cases.
For brevity, the energies of the discrete states are indicated by
$\varepsilon_{n,j}=E_{n,0}+\mu_j$ in the following.
We stress that the transverse energies $E_n$ are fixed, since the
cross-section is constant, although rotated. Contrary, the position of
$\varepsilon_{n,j}$ depends upon the twist, as we will analyze in
detail in Section~\ref{sec4}. In fact, they are the resonant energies
that correspond to a local maximum (Breit-Wigner) or a zero (Fano) in
the transmission spectra.

\section{Numerical approach}
\label{sec3}

To obtain the transmission amplitudes of the twisted QW we solve the
3D Schr\"odinger equation with open boundaries through the quantum
transmitting boundary method~\cite{lent90} (QTBM).  Electrons
are injected from the left lead (see Fig.~\ref{fig1}) in a given
transverse mode, and can be either reflected or transmitted to the
right lead.  With this boundary condition, the differential
equation of motion is solved in the internal points of the domain,
leading to complex transmission/reflection amplitudes for every
channel of the right/left leads.
We adopt a curved coordinate system naturally defined by the twist
function of Eq.~(\ref{eq5}) $\mathbf{r}=(x,y,z) \to \mathbf
{r'}=(x',y',z')$.  This new coordinate system follows the QW twist and
``sees'' a straight QW.  However, also the equation of motion must be
transformed according to the $\mathbf{r} \to \mathbf{r'}$ relation.
In order to do so, we need the metric tensor of the system
$\mathbf{G}(\mathbf{r})$ with components $G_{ij}(\mathbf{r}) =
(\partial_{i}\mathbf{r'}) \cdot (\partial_{j}\mathbf{r'})$, together
with its inverse $\mathbf{G}^{-1}$ with components $G^{ij}$. Here, we
used the definitions $\partial_{i}=\partial / \partial x^{i}$ and
$(x,y,z)=(x^{1},x^{2},x^{3})$.  In the curved coordinate system the
Hamiltonian reads~\cite{dacosta82,ferrari08}:
\begin{eqnarray}
&&\mathcal{H}(\mathbf{r}) = -\frac{\hbar^{2}}{2m}
\sum_{i,j=1}^{3}\frac{\partial_{i}}{\sqrt{G}}
\left(\sqrt{G}G^{ij}\partial_{j} \right)              \nonumber \\
&& = -\frac{\hbar^{2}}{2m}
\sum_{i,j=1}^{3}\left[G^{ij}\partial^{2}_{ij}
- \left( \sum_{k,l=1}^{3} G^{kl}\partial^{2}_{kl}
\mathbf{r'}\cdot G^{ij}\partial_{i}\mathbf{r'} \right)
\partial_{j}   \right] ,                              \nonumber
\end{eqnarray}
where $G>0$ is the determinant of $\mathbf{G}$ and
$\partial^{2}_{ij}=\partial_{i}\partial_{j}$.  Now it is easy to
define a rectangular mesh following the QW in the new coordinate
system, and discretize $\mathcal{H}$ through a finite-difference
scheme. The corresponding Schr\"odinger equation
$\mathcal{H}\psi=E\psi$ is then solved, with open boundary conditions
at the two edges of the QW, as described above.  The QTBM takes as an
input the kinetic energy $E-E_n>0$ of the incoming electron and the
wave function of the transverse mode $n$, and gives as an output the
transmission/reflection amplitudes in the different
channels~\cite{cuoghi09}.  For this reason, a new calculation must be
performed for every $E$ in a chosen set over the range of interest,
with $E>E_n$.

Actually, in order to find the resonances we also used a
complementary technique: the complex scaling approach described in
Appendix~\cite{kovarik07}.  This method leads to a complex-eigenvalue
problem that allows one to identify in a straightforward way the
resonances originated by the bound states of $V$.  Moreover, it gives
the energy levels of bound states below $E_1$, not achievable with the
QTBM.  In fact, the QTBM gives the transmission amplitude of the
different transverse modes as a function of the carrier energy, and
the position of transport resonances must be detected subsequently, as
a relevant peak or dip in the transmission probability and as a
continuum (abrupt) phase shift for a Breit-Wigner (Fano)
resonance. However, the complex scaling method results to be very
demanding from the computational point of view and we used its results
only as a reference for specific cases. We leave the comparison of the
two methods to a subsequent work.

\section{Transmission spectra}
\label{sec4}

As anticipated in Section~\ref{sec2} the transmission spectra of the
straight QW can be obtained from a one-dimensional equation of motion
with the potential $V$.  In order to mix transmission channels, the
transverse/longitudinal separability must be lifted.  However, a
generic deformation of the QW section along the wire, not only couples
different transverse modes, but also introduces additional resonant
energies, as in the case of a closed cavity attached at a
side~\cite{price92,porod92}.  This can make difficult to expose the sole
effect of the coupling between the continuum and discrete spectra.
For this reason, as well
as for technological relevance, we choose a kind of deformation that
does not alter the shape of the QW cross section, but only its
orientation, and does not introduce further resonances.  In fact, the
QW twist has only two effects: first, it couples the transmission
channels so that the transmission probability for a carrier injected in a given
transverse mode also has traces of quasi-bound states of different
modes; second, it increases the energies of quasi-bound states.
\begin{figure}
\includegraphics[width=0.9\linewidth]{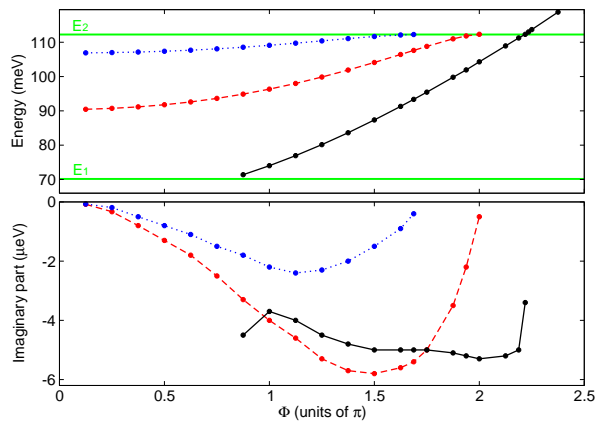}
\caption{
Top panel: position of transmission resonances in the ground-mode
spectrum as a function of the twist angle $\Phi$, for $\nu=2.95$.  The
resonant energies correspond to the real part of the complex-scaled
eigenvalues.  The threshold energies of the ground and first-excited
transverse modes are indicated by the two horizontal lines labeled as
$E_1$ and $E_2$, respectively. The three curves correspond to the
quasi-bound states $\epsilon_{2,3}$ (dotted line), $\epsilon_{2,2}$
(dashed line), $\epsilon_{2,1}$ (solid line). Bottom panel: Imaginary
part of the complex-scaled eigenvalues, corresponding to the half
width of the resonance peaks or dips. While the two Fano resonances
$\epsilon_{2,3}$ and $\epsilon_{2,2}$ (see also Fig.~\ref{fig4})
disappear as they reach $E_2$, $\epsilon_{2,1}$ is still present in
the spectrum at energies exceeding $E_2$ in the form of a broad and
shallow dip (see also Fig.~\ref{fig5}).  }\label{fig3}
\end{figure}
These effects can be seen from the two panels of
Fig.~\ref{fig3}, where we report the position (top panel) and width
(bottom panel) of the resonances in the transmission spectrum of the
ground transverse channel, as a function of the twist angle.  In
particular, for each angle $\Phi$ we inject a carrier in the ground
transverse mode (with energy $E_1$) and with several longitudinal
kinetic energies, from zero to $E_2-E_1$.  From the curves of
transmission amplitude vs. total energy $E$ (see e.g. Figs.~\ref{fig4} and
\ref{fig5}) we determine the position of the resonances and
obtain the imaginary part of the eigenvalue as $-\Gamma/2$, where
$\Gamma$ is the peak or dip width.  Note that the above complex
eigenvalues correspond to quasi-bound states with a mean lifetime
$\hbar/\Gamma$ of the order of $100$~ps.

By comparing Fig.~\ref{fig3} with the levels of the straight QW given
in Fig.~\ref{fig2}, the origin of the two resonances at higher energy
(dotted and dashed lines in Fig.~\ref{fig3}) is clear.  In fact, for
$\Phi\sim 0$ only two quasi-bound levels lie between $E_1$ and $E_2$,
namely $\epsilon_{2,2}$ and $\epsilon_{2,3}$.  They both belong to
channel $2$ so that at zero twist, they do not appear in the
transmission spectrum, as it can be gathered from the vanishing
imaginary part of their eigenvalue in the bottom panel of
Fig.~\ref{fig3}.  As the twist is introduced, the two levels above
appear as slightly asymmetric Fano dips in the transmission
probability, as reported in the top panels of Fig.~\ref{fig4} for the
case of $\Phi=\pi/2$.  However, the Fano character of the resonances
is better revealed by the abrupt jump of $\pi$ in the transmission
phase $\theta$, as shown in the bottom panels of Fig.~\ref{fig4}.
\begin{figure}
\includegraphics[width=0.9\linewidth]{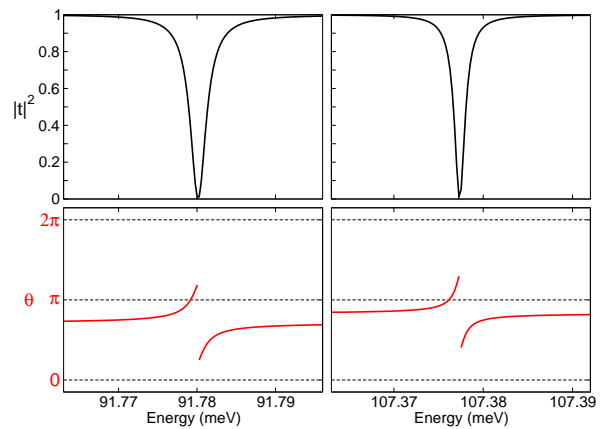}
\caption{ Ground-mode transmission probability (upper panels)
and phase (lower panels) around the resonances $\epsilon_{2,2}$ (left
panels) and $\epsilon_{2,3}$ (right panels) for a twist angle
$\Phi=\pi/2$. In both cases, the phase $\theta$ shows the typical
$\pi$ discontinuity of Fano resonances.  }\label{fig4}
\end{figure}
The third resonance of Fig.~\ref{fig3} (solid line) appears around
$\Phi\simeq 0.85 \pi$ from the low-energy threshold $E_1$.  It is
again a Fano resonance, as can be gathered from the left panels of
Fig.~\ref{fig5}, showing the transmission probability and phase at
$\Phi=\pi$. This is confirmed by results of the complex-scaling
approach, ascribing the resonance to the quasi-bound level
$\epsilon_{2,1}$.
\begin{figure}
\includegraphics[width=0.9\linewidth]{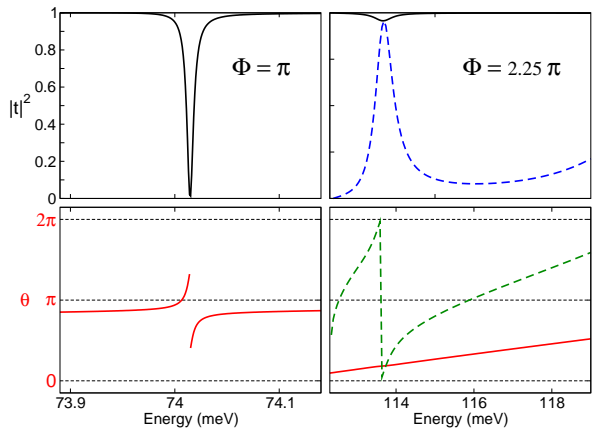}
\caption{ Ground-mode transmission probability (upper panels)
and phase (lower panels) around the resonance $\epsilon_{2,1}$ for two
different twist angles, namely $\Phi=\pi$ (left panels) and
$\Phi=2.25\pi$ (right panels, solid line).  In the first case
$\epsilon_{2,1}<E_2$, and a sharp Fano dip is found. In the second
case $\epsilon_{2,1}>E_2$, and its position is only indicated by a
very shallow dip. In the latter case, also the transmission spectrum
of the first-excited mode is reported (dashed line), revealing a
pronounced Breit-Wigner resonance.  
}\label{fig5}
\end{figure}
In fact, as the twist increases from $0$ to $\pi$, both levels
$\epsilon_{2,1}$ and $\epsilon_{1,3}$ reach the threshold $E_1$. However,
while the former shows up as a resonance in the transmission spectrum,
the latter disappears as it enters the traveling-states region, with its
imaginary part going to zero.  Again, this behavior can be traced by the
complex scaling method alone, since the lower energy accessible with
the QTBM is $E_1$.
When $\epsilon_{2,1}$ enters the energy range of first-mode traveling
states, it appears as a Fano resonance since it is a second-mode
quasi-bound state.  Moreover, contrary to the other two resonances, it
appears with a significant width (bottom panel of Fig.~\ref{fig3})
from the beginning, since at $\Phi\simeq 0.85 \pi$ the coupling
between the modes is already strong.

When the twist increases from $\pi$ to $2\pi$ the three quasi-bound
levels described above also increase their energy. However, as
$\epsilon_{2,2}$ and $\epsilon_{2,3}$ approach the threshold of the
second channel, their width decreases and they finally disappear, with
the width going to zero, when their energy reaches $E_2$. The behavior
of $\epsilon_{2,1}$ is different.  In fact, its resonance width is
always of the order of $4 \mu eV$ until the energy reaches $E_2$,
where the width increases by order of magnitudes.  This is
shown in the right panels of Fig.~\ref{fig5}, where the transmission
probability (top) and phase (bottom) are shown, for a twist
$\Phi=2.25\pi$. Here, $\epsilon_{2,1}>E_2$, and the second
transmission channel becomes available. The solid line is the
first-channel to first-channel transmission, and shows a tiny dip at
the quasi-bound state position, reminiscent of the prominent Fano dip
of the single-channel case.  
The dashed line is the second-channel to
second-channel transmission, showing a clear Breit-Wigner resonance
with the corresponding continuous phase lapse of $\pi$.~\cite{nota1}
%
%
This is not surprising, since in this case the quasi-bound state has
the same transverse mode of the transmission channel.

\begin{figure}
\includegraphics[width=0.9\linewidth]{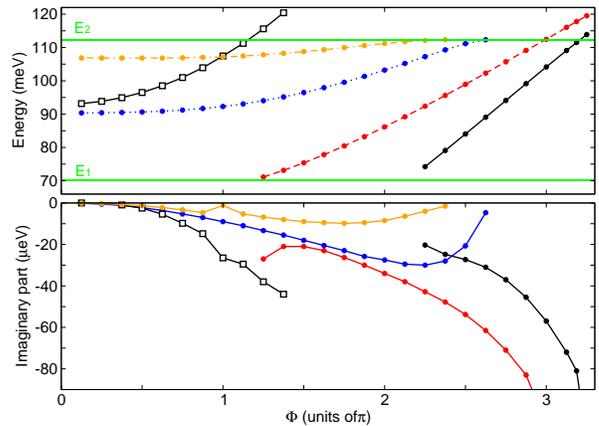}
\caption{
Same as Fig.~\ref{fig3}, but for a deeper potential well, with
$\nu=3.95$.  Here, four quasi-bound states with a first-excited
transverse mode are present: $\epsilon_{2,1}$ (solid line, with filled
circles), $\epsilon_{2,2}$ (dashed line), $\epsilon_{2,3}$ (dotted
line), $\epsilon_{2,4}$ (dot-dashed line), together with a quasi-bound
state with a second-excited transverse mode: $\epsilon_{3,1}$ (solid
line with empty squares). The two resonances $\epsilon_{3,1}$ and
$\epsilon_{2,4}$, with different transverse modes, cross around
$\Phi=\pi$ with a repulsion of their imaginary components.  All the
resonances in the $[E_1,E_2]$ range are of Fano type.  
}\label{fig6}
\end{figure}

\begin{figure}
\includegraphics[width=0.9\linewidth]{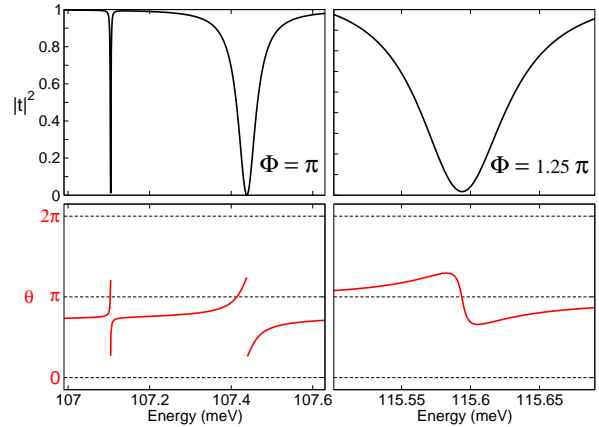}
\caption{
Ground-mode transmission probability (upper panels) and phase (lower
panels) showing the resonance $\epsilon_{3,1}$ just after the
crossing with $\epsilon_{2,4}$ (at $\Phi=\pi$, left panels) and above
the second mode threshold $E_2$ (at $\Phi=1.25\pi$, right panels).  In
the first case the $\pi$ discontinuities of the transmission phase
evidence the Fano character of the resonances. In the second case a
third channel is available, and the resonance does not present either
Fano or Breit-Wigner character.  
}\label{fig7}
\end{figure}

The case with $\nu=3.95$, with a deeper potential well $V$ in the
twisted region, presents additional effects. In fact, at zero twist
the energy range between $E_1$ and $E_2$, where only the ground
channel is open, contains bound states of two different transverse
modes, namely the first excited ($\epsilon_{2,3}$ and
$\epsilon_{2,4}$) and the second excited ($\epsilon_{3,1}$), as shown
in Fig.~\ref{fig2}.  As the wire is twisted, the energy of the above
three quasi-bound states increases, as illustrated in the top panel of
Fig.~\ref{fig6}.  
However, the level $\epsilon_{3,1}$ (solid line with empty squares)
increases faster than the other two, it crosses $\epsilon_{2,4}$
around $\Phi=\pi$ and goes beyond $E_2$.  First of all we note again
that, in the $[E_1,E_2]$ range, the transport resonances corresponding
to the above quasi-bound states are Fano resonances, in agreement with
the fact that they originate from a transverse mode different from
that of the transport channel.  This is shown in the left panels of
Fig.~\ref{fig7}, reporting the transmission probability and phase of
the ground channel at $\Phi=\pi$, just after the crossing.  At the
crossing, we also find a {\it repulsion} of the imaginary component of
the eigenvalues, visible in the bottom panel of Fig.~\ref{fig6}.

When $\epsilon_{3,1}>E_2$, i.e. it enters the energy region with two
transport channels, it does not disappear, as $\epsilon_{2,3}$ and
$\epsilon_{2,4}$ do at larger twist, but simply changes its
characters. Now the minimum of the dip does not reach zero, and the
phase does not present the $\pi$ discontinuity (Fig.~\ref{fig7}, right
panels). Here in fact, a third transmission channel is available, this
lifting the rigid zero-transmission properties of a two-channel case
Fano resonance.

\begin{figure}
\includegraphics[width=0.9\linewidth]{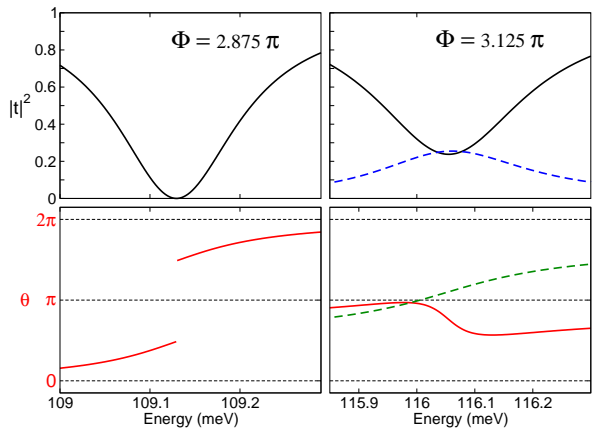}
\caption{
Transmission probability (upper panels)
and phase (lower panels) around the resonance $\epsilon_{2,2}$ for 
the system with $\nu=3.95$.
Two different twist angles are considered, with the resonance
close to the second-channel threshold: $\Phi=2.875\pi$ (left panels),
where $\epsilon_{2,2}<E_2$ and $\Phi=3.125\pi$ (right panels), where
$\epsilon_{2,2}>E_2$. In the second case, both the ground mode
(solid line) and first-excited mode (dashed line) transmission probabilities
are reported.
As for the case of Fig.~\ref{fig7}, after the
crossing of $E_2$ the $\pi$ discontinuity of the
phase is lost, due to the availability of a third transmission channel.
The broad peak in the first-excited channel is a Breit-Wigner resonance.
  }\label{fig8}
\end{figure}

As already mentioned, the two resonances $\epsilon_{2,3}$ and
$\epsilon_{2,4}$, present in the spectrum since small twist angles,
disappear as they reach $E_2$, with their width going to zero.  Two
additional resonances enter the ground-mode region at larger twists:
$\epsilon_{2,2}$ and $\epsilon_{2,1}$, represented in Fig.~\ref{fig6}
by a dashed line and a solid line, respectively. They are also Fano
resonances, but after reaching $E_2$ they do not vanish. In fact,
their width increases and their minimum does not reach zero, as shown
in Fig.~\ref{fig8} for $\epsilon_{2,2}$. Obviously, in this region
they are also present in the transmission spectrum of the
first-excited channel as Breit-Wigner resonances (dashed line in
Fig.~\ref{fig8}), since their transverse mode is the first-excited one
as well. Correspondingly, their transmission phase presents a smooth
evolution of $\pi$.

\section{Conclusions}
\label{sec5}

By solving the open-boundary Schr\"odinger equation through the QTBM
we obtained the transmission spectra of the twisted QW.  The effect of
the twist can be summarized in the following points. 
\emph{First}, the twist is able to mix different transmission channels
in spite of the fact that the transverse QW section is not altered
(but only rotated). Furthermore, bound states of $V$ are coupled to
traveling states and appear as resonance peaks/dips in the transmission
spectra. No additional resonances are introduced.
\emph{Second}, the character of the resonance depends upon the transverse
mode of the original bound state.
In fact, when the latter is equal to the transverse mode of the transmission
channel, we find a Breit-Wigner resonance, otherwise we find a Fano resonance.
In case more than two channels are available, we do not find the
$\pi$ discontinuity of the transmission phase typical of Fano
resonances.
\emph{Third}, the twist increases the energy of quasi-bound levels.
The higher the transverse mode of the quasi-bound state, the faster
its energy is increased. However, the change of resonance width is
non-monotonic with the twist. In general, it increases from zero 
when the energy of the bound state is already in the transport
region in the straight QW, and decreases as the above energy reaches
the threshold of the transmission channel with the same transverse
mode as the quasi-bound state.
\emph{Forth}, resonances that are present from the beginning in the
first-channel region disappear as they reach $E_2$, while resonances
that enter the $E_1$-$E_2$ region at a finite twist, persist in the
multi-channel region.
The strict behavior described above could help in anticipating the
characters of transmission spectra of QW locally twisted once the
spectra of the straight wire is known. Furthermore, it supports the
idea that the twist can reduce the effects of localized states on
quantum conductance, since it shifts their levels towards higher
energies, possibly beyond Fermi level of the quasi-1D nanostructure.

\begin{acknowledgments}
We thank P.~Bordone, G.~Ferrari and H.~Kova\v{r}\'{i}k
for most helpful discussions.
\end{acknowledgments}

\section*{APPENDIX: COMPLEX SCALING METHOD}

So far we have identified resonant energies with singular points of
the reflection/transmission coefficient. In the framework of the
complex scaling method, introduced in the '70 by Aguilar, Baslev and
Combes \cite{aguilar71, balslev71} (for a review see also
Ref.~\onlinecite{CFKS}), resonances are identified with the complex
eigenvalues of a non-symmetric linear operator obtained from the
original one by analytic complex deformation. The real part of such
complex-valued eigenvalues coincides with the usual resonance energy
level, while the imaginary part is associated with the resonant state
lifetime. The complex scaling method has been employed in
Ref.~\onlinecite{kovarik07} to twisted QW in order to prove
the existence of resonances and here we briefly resume it.

Let $\omega = \left [- \frac 12 L_y , + \frac 12 L_y \right ] \times
\left [- \frac 12 L_z , + \frac 12 L_z \right ] \subset R^2$ be the
rectangular cross section of our QW.  For a given $x\in R $ and
$(y,z)\in\omega$ we define the mapping given in Eq.~(\ref{eq5})
where $\varphi (x) = \epsilon \alpha (x)$ and where $\alpha:R \to R $
is a differentiable function which represent the twisting and
${\epsilon}\ge 0$ is a real-valued parameter which represents the
strength of the twisting.  Let $\Omega_{{\epsilon}}$ be the twisted
QW and let
\begin{eqnarray*}
{H}_{\epsilon} = - \frac {\hbar^2}{2m} \Delta +V(x) 
\end{eqnarray*}
be the time independent Schr{\"o}dinger operator on
$\Omega_{{\epsilon}}$, that is the wave function $\psi$ belongs to
$L^2(\Omega_{{\epsilon}})$ with Dirichlet boundary conditions at
$\partial\Omega_{{\epsilon}}$.  $V$ represents the external potential
Eq.~(\ref{eq2}), depending only on the longitudinal variable $x$.  In
the following, for the sake of definiteness, let us assume the units
choice such that $\frac {\hbar^2}{2m} =1$.

In order to analyze the operator ${H}_{\epsilon}$ we go back to the
untwisted tube $\Omega$.  The operator $H_\epsilon$ then takes the
form $K_\epsilon$
\begin{eqnarray*}
K_{\epsilon} = - \partial_{yy}^2- \partial_{zz}^2-
[\partial_x+{\epsilon} \alpha' (x)\, \partial_{\tau}]^2+ V(x) =K_0 +
U_{\epsilon} \, ,
\end{eqnarray*}
where
\begin{eqnarray*}
\partial_\tau = y \partial_z - z \partial_y
\end{eqnarray*}
and
\begin{eqnarray*}
K_0  = -  \partial_{xx}^2 -  \partial_{yy}^2- \partial_{zz}^2 +V(x)
\end{eqnarray*}
and
\begin{eqnarray*}
U_{\epsilon} &=& -  [\partial_x+{\epsilon} \alpha' (x)\, \partial_{\tau}]^2 
+ \partial_{xx}^2 \\ 
&=&  - 2{\epsilon}\,  \alpha' ( x) \partial^2_{x\tau} 
- {\epsilon}\, \alpha'' ( x)\, \partial_{\tau} -{\epsilon}^2 
\left[ \alpha ' ( x)\right ]^2\, \partial_{\tau \tau}^2 \, . 
\end{eqnarray*}

The operator $K_{\epsilon}$ is a symmetric operator on $L^2 (\Omega)$
with Dirichlet boundary conditions at $\partial \Omega$.  The
spectrum of $K_0$ is given by Eq.~(\ref{eq4}),
that is, the spectrum of $K_0$ admits
embedded eigenvalues in the continuous spectrum.  In
Ref.~\onlinecite{kovarik07} it has been proved that such embedded
eigenvalues become resonances when we add the perturbation
$U_\epsilon$ to $K_0$.  Resonances are defined by employing the
method of exterior complex scaling to the operator $K_{\epsilon}$,
provided that the potential $V$ is a bounded potential which extends
to an analytic function with respect to $x$ in some sector, and the
twisting function $\alpha (x)$ extends to analytic function with
respect to $x$ in a suitable complex set.  The exterior complex
scaling method consists in introducing the mapping $S_\theta$, which
acts as a complex dilation in the longitudinal variable $x$:
\begin{eqnarray*}
(S_{\theta}\psi)(x,y,z) = e^{\theta/2}\psi(e^{\theta} x,y,z)\, , \quad
\theta \in C , \, \Im \theta >0 .
\end{eqnarray*}
The transformed operator is not a symmetric operator and it takes the form
\begin{eqnarray*}
K_{\epsilon} (\theta ) = S_{\theta} K_{\epsilon} S^{-1}_{\theta} 
= K_0 (\theta ) + U_{\epsilon} (\theta)\, ,
\end{eqnarray*}
where
\begin{eqnarray*}
K_0 (\theta ) = S_{\theta} K_{0} S^{-1}_{\theta} = -e^{-2\theta} \, 
\partial_{xx}^2 - \partial_{yy}^2-\partial_{zz}^2 +V(e^{\theta}x) \,
\end{eqnarray*}
and
\begin{eqnarray*}
U_{\epsilon} (\theta ) &=& S_{\theta} U_{{\epsilon}} S^{-1}_{\theta} 
\\
&=& - 2{\epsilon}\, e^{-\theta}\, \alpha'( e^\theta x) \partial^2_{x \tau} 
- {\epsilon}\, \alpha'' ( e^\theta x)\, \partial_{\tau} 
\\
&& -{\epsilon}^2 
\left[ \alpha'( e^\theta x)\right ]^2 \, \partial_{\tau \tau}^2 \, .
\end{eqnarray*}

\begin{figure}[h]
\begin{center}
\includegraphics[height=2cm,width=8cm]{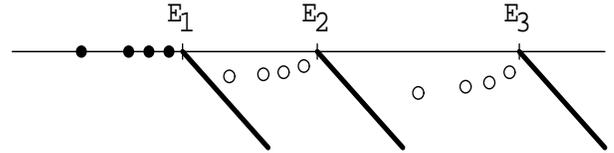}
\caption{The essential spectrum of $K_{\epsilon} (\theta )$ is given
by the half-lines $E_{n} + e^{-2i\Im \theta } R^+$ (full
lines). The eigenvalues of $K_{\epsilon} (\theta )$ (denoted by open
circle) with strictly negative imaginary part are the 
\emph{resonances} of $H_{\epsilon}$; for energies below the threshold 
$E_{1}$ the eigenvalues of $K_{\epsilon} (\theta )$
(denoted by full circle) are purely real valued and they are
eigenvalues of $H_\epsilon$.}  
\label {fig9} 
\end{center} 
\end{figure}

Then, the essential spectrum of $K_{\epsilon}(\theta )$ consists of
the sequence of the half-lines (Fig.~\ref{fig9}) $E_{n} +
e^{-2i\Im \theta } R ^+ $, $n =1,2,\ldots $, and, by a
standard argument, it turns out that the eigenvalues of $K_{\epsilon}
(\theta )$ are analytic functions of $\theta$, they are in fact
independent of $\theta$. These non-real eigenvalues of $K_{\epsilon}
(\theta )$, for $\theta$ such that $\Im \theta >0$, are identified
with the resonances of $K_{\epsilon}$ (and hence with the resonances
of $H_\epsilon$) \cite{CFKS}.  


\begin{thebibliography}{25}
\bibitem [{not()}]{nota1}%
  \BibitemOpen
  \href@noop {} {}\bibinfo {note} {Note that in Fig.~\ref{fig5}, the abscissa
  of the right panels has a scale much larger than in the left panels or in
  Fig.~\ref{fig4}. As a consequence, the peaks/dip here reported have a width
  much larger than the previous case, as described in the text. Also note that
  the transmission phase $\theta$ has been reduced to the $[0,2\pi]$ region so
  that a value of $2\pi+a$ is rescaled and reported as $a$. The vertical dashed
  line is not a phase jump: on the contrary, it has been introduced to indicate
  a continuity of the curve, as opposed to the Fano case of the left
  panel.}\BibitemShut {Stop}%
\end{thebibliography}

\providecommand{\noopsort}[1]{}\providecommand{\singleletter}[1]{#1}%

\end{document}